\documentstyle[aps,aps12]{revtex}
\begin{document}   
\titlepage

\title{On the stability of renormalizable expansions \\in
three-dimensional gravity}

\author
{Shun'ya MIZOGUCHI
\footnote{
e-mail: mizoguch@x4u2.desy.de}
}
\address
{II. Institut f\"{u}r Theoretische Physik, 
            Universit\"{a}t Hamburg\\
            Luruper Chaussee 149, 22761 Hamburg, F.R.G.
}

\author
{Hisashi YAMAMOTO
\footnote{e-mail: hisashi@yisun1.yukawa.kyoto-u.ac.jp}
}
\address
{Department of Food Studies, Faculty of Home Economics, \\
Doshisha Women's College of Liberal Arts, Kyoto 602, Japan
}
\maketitle

\vspace{.5cm}

\abstract{
Preliminary investigations are made for the stability 
of the $1/N$ expansion in three-dimensional gravity 
coupled to various matter fields, which are power-counting 
renormalizable. For unitary matters, 
a tachyonic pole appears in the spin-2 part of the leading
graviton propagator, which implies the unstable flat space-time, 
unless the higher-derivative terms are introduced.
As another possibility to avoid this spin-2 tachyon, we 
propose Einstein gravity coupled to non-unitary 
matters. 
It turns out that a tachyon appears in the spin-0 or -1 
part for any linear gauges in this case, but it can be removed 
if non-minimally coupled scalars are included.
We suggest an interesting model which may be stable and possess 
an ultraviolet fixed point.
}
\newpage

\section{Introduction}

\indent

One of the important issues in modern particle physics is to clarify
quantum properties of gravitation. Although most of the recent 
interest has been focused on two dimensions, efforts have also 
been made toward constructing quantum gravity in higher 
dimensions, as statistical models of space-time and matters. 
Among them, numerical simulations of dynamical random
lattices have given evidences of phase transitions in three 
and four dimensions \cite{DTL}. The renormalization-group study 
of $2+\epsilon$-dimensional gravity has revealed interesting 
phenomena, suggestive of higher-dimensional properties \cite{KKN}.
However, apart from the formulation as a Chern-Simons 
gauge theory \cite{CSW} and the canonical formalism \cite{Ash}, 
conventional analytical studies of three- and four-dimensional 
Einstein gravity coupled to matters have to face the problem 
of non-renormalizability \cite{Weinberg}.

Perturbative non-renormalizability does not imply 
the non-existence of continuum field theory, if regarded 
as a low-energy effective theory of some fundamental one, such 
as superstring. It may be even important to extract universal 
low-energy properties of quantum gravity from such theories.
To embody this program in the continuum approach, however, the
realization of renormalizability is technically necessary for
any actual prediction.To consider this problem, it is useful here 
to learn from other examples of non-renormalizable field theories. 
It is well known that there exist a class of weak-coupling
non-renormalizable field theories which are rendered 
renormalizable by using resummation methods such as the $1/N$ 
expansion. We may list, for example, four-fermi  
and non-linear sigma models in three dimensions 
\cite{4fermi,NLSM}, in which ultraviolet divergences are 
softened by use of dressed propagators for auxiliary fields.
Moreover, with no restriction to coupling strength, this 
``non-perturbative'' expansion yields reliable predictions 
such as of phase transitions and associated exponents even for 
a finite $N$ \cite{4fermi,NLSM,MSHKK}. The application of such 
a method to gravity system is then appealing as an approach to 
explore quantum properties of gravitation.

The idea of resummation in quantum gravity is not new.  
Tomboulis first applied the $1/N$ expansion to four-dimensional
gravity coupled to massless spinors, where $N$ is the number of
fermion species \cite{T}.  
With the inclusion of higher-derivative counterterms, 
renormalization properties and behaviors of graviton 
propagator were examined in the large-$N$ limit. 
Recently Kugo applied this expansion to three-dimensional Einstein
gravity coupled to massive scalar fields \cite{Kugo}.  
In three dimensions  
Einstein gravity coupled to matters were shown to be formally 
(power-counting) renormalizable via the $1/N$ expansion without 
introducing higher-derivative terms, although
the graviton propagator turned out to develop a tachyonic pole.
After this work, no further investigation has been made along this
approach.

To examine quantum nature of space-time, one must handle two kinds 
of quantum effects. 
One is from a self-interaction of gravity, and another 
is from quantum fluctuations of matters turning back into gravity 
through a matter-gravity coupling (back reaction).
Roughly speaking, the present approach ($1/N$ expansion) 
is such a method that the second effect is taken into account 
to leading order, while the first one is incorporated 
with it in higher orders.
This method may be regarded as complementary to the usual approach,   
in which the first effect is taken into account from the first. 
For example, in modern numerical works of quantum gravity
one usually first sums over all simplicially decomposed 
configurations 
of space-time to explore the phase structure of pure-gravity system,
and next takes a few kinds of matters into simulations to see
whether their effects are small or not.
In other words, the analyses are performed in the regime of
pure gravity and its neighborhood ({\em gravity dominant regime}).
To the contrary, the $1/N$ expansion gives by definition a priority 
to matter fluctuations, and hence formally its prediction is 
relevant in the regime of many matters 
({\em matter dominant regime})
although the actual applicability limit is not known a priori.
We believe that the studies from both directions are useful 
for the full understanding of quantum gravity with couplings 
to general matters.

In this paper, we will report our preliminary
results for the stability of renormalizable expansions in
three-dimensional gravity coupled to various matter fields,
as a first step toward constructing stable (tachyon-free) 
theories.  
In sect.II we first review the result of ref.\cite{Kugo} with brief 
discussions added.
In sect.III we consider couplings to various unitary matter 
fields other than minimal scalars,
such as spinors, U(1) gauge fields and non-minimal scalars, 
and examine whether we can have a tachyon-free graviton propagator
for such unitary matters.
We will see that in all cases above tachyonic poles 
show up in the leading-order graviton propagators. 
In these analyses we will find that the spin-2 part of matter 
one-loop corrections are quantized 
and proportional to their physical degrees
of freedom, the situation being similar to that of conformal 
anomaly in two dimensions.    
In sect.IV we will then propose two independent
possibilities to circumvent this spin-2 tachyon.
The first one is to add higher-derivative terms for gravity. 
The second is to couple Einstein gravity to {\em non-unitary} matters. 
In the latter case, although the spin-2 part of the 
graviton propagator is free from a tachyon, it turns out 
to appear in the {\em gauge-dependent (spin-0,1) part}.  
In sect.V we will prove that this new tachyonic pole is 
inevitable for any choice of linear gauges, and indicate a class 
of tachyon-free models by including non-minimal scalars.
Sect.VI is devoted to summarize our results, and to give the future 
prospects expected from them.  
In appendix A is listed the notation of the
tensors which appear in the graviton propagator.
Appendix B includes some technical
formulas for the tensor calculus. The Feynman rules for the
calculations of matter one-loop vacuum polarizations are 
summarized in Table I.  

        We employ in this paper the conventions 
of the flat Minkowski metric  
\begin{equation}
\eta^{\mu\nu}={\rm diag}(1,-1,-1)
\end{equation}
and the Ricci tensor 
\begin{equation}
R_{\mu\nu}=R^{\lambda}_{~\mu\lambda\nu}
=\partial_{\nu}\Gamma^{\lambda}_{~\mu\lambda}
-\partial_{\lambda}\Gamma^{\lambda}_{~\mu\nu}
+\Gamma^{\lambda}_{~\tau\nu}\Gamma^{\tau}_{~\mu\lambda}
-\Gamma^{\lambda}_{~\tau\lambda}\Gamma^{\tau}_{~\mu\nu}.
\end{equation}

\section{$1/N$ expansion}
\subsection{Gravity coupled to scalar fields}
\indent

        To present the prototype of the $1/N$ expansion in
three-dimensional gravity, we review here the case of the minimal
coupling to $N$ massive scalar fields 
$\phi_i$ $(i=1,\ldots,N)$ \cite{Kugo}. 
The lagrangian of the system is
given by 
\begin{equation}
{\cal L}= \frac{1}{\kappa^2}\sqrt{g}R~+
~{\cal L}_{\rm{matter}}~+
~{\cal L}_{\rm{FP+GF}},
\label{eq:totalL}
\end{equation}
\begin{equation}
{\cal L}_{\rm{matter}}
={\cal L}_{\rm{scalar}}\equiv \frac{1}{2}\sqrt{g}
\left(g^{\mu\nu}\partial_{\mu}\phi_i 
\partial_{\nu}\phi_i
-m^2\phi_i^2\right),
\end{equation} 
where 
${\cal L}_{\rm{FP+GF}}$ is the gauge-fixing and the FP ghost 
terms associated with general coordinate gauge symmetry. 
The construction of the $1/N$ expansion in this system is most
conveniently achieved in a diagrammatic way with the following
non-local action:
\begin{eqnarray}
\Gamma&=&N
\left[
\int d^3x\left( \frac{1}{\kappa^2}\sqrt{g}R+
 \frac{1}{2}\sqrt{g}
\left(g^{\mu\nu}\partial_{\mu}
\phi_i\partial_{\nu}\phi_i-m^2\phi_i^2\right)+
{\cal L}_{\rm{FP+GF}}\right)\right.\nonumber\\
&&\left.
~~~+  ~\frac{i}{2}\log{\rm Det}(-\Delta(g_{\mu\nu},m))+\int d^3x
{\cal L}^{(0)}_{\rm{count}}
\right],
\label{eq:starting}
\end{eqnarray}
where 
$\Delta(g_{\mu\nu},m)
=\partial_{\mu}\sqrt{g}g^{\mu\nu}\partial_{\nu}
+\sqrt{g}m^2$. 
${\cal
L}^{(0)}_{\rm{count}}$ 
denotes the
counterterms for renormalization. 
In (\ref{eq:starting}) we have rescaled $\kappa^2$ and $\phi$ as 
$\kappa^2\longrightarrow N^{-1}\kappa^2$ and 
$\phi\longrightarrow N^{\frac{1}{2}}\phi$, respectively. 
The ghosts and the gauge parameters are rescaled similarly. 
The loop expansion based on the Feynman rules read off from
(\ref{eq:starting}) provides the expansion of the full effective
action in a power series of $1/N$ instead of $\kappa^2$.
To this rules must be added the provision 
that when one calculates to higher orders in $1/N$, 
one omits the closed scalar-loop graphs, since they
are already taken into account in the leading-order graviton
propagator and vertices. Indeed, the leading 
graviton propagator already includes the contributions from an 
arbitrary number of scalar one-loop self-energy insertions, 
and the leading vertices also contain one-loop corrections (fig.1). 
Thus, in essence, the $1/N$ expansion is a gauge-invariant 
rearrangement of the Feynman diagrams.

        We now calculate the leading graviton propagator. 
It is convenient to define the fluctuation around the flat metric
as follows
\begin{eqnarray}
\widetilde{g}^{\mu\nu}&\equiv&\sqrt{g}{g}^{\mu\nu}\nonumber\\
&=&\eta^{\mu\nu}+\kappa\widetilde{h}^{\mu\nu},   \label{eq:flat}
\end{eqnarray} 
and 
\begin{eqnarray}
{g}_{\mu\nu}
&=&\eta_{\mu\nu}+\kappa h_{\mu\nu}.  \label{eq:flat1}
\end{eqnarray}
$\widetilde{h}^{\mu\nu}$ and $h^{\mu\nu}$ are related with each other by 
\begin{eqnarray}
\widetilde{h}^{\mu\nu}
&=& \frac{1}{2}\eta^{\mu\nu}h-h^{\mu\nu}
+{\rm O}(\kappa),~~~
\\
h^{\mu\nu}&=&\eta^{\mu\nu}\widetilde{h}-\widetilde{h}^{\mu\nu}
+{\rm O}(\kappa),~~~
\label{eq:h->th}
\end{eqnarray}
where $h=h^{\mu}_{\mu}$ and $\widetilde{h}=\widetilde{h}^{\mu}_{\mu}$. 
Their indices are raised and lowered 
by the flat metric $\eta_{\mu\nu}$. 
In the following we will take $\widetilde{h}^{\mu\nu}$ as our basic quantum
field around the flat space-time. 
In terms of $\widetilde{h}^{\mu\nu}$, ${\cal L}_{\rm{scalar}}$ is expanded as 
\begin{eqnarray}
{\cal L}_{\rm{scalar}}
&=& \frac{1}{2}\left(
\eta^{\mu\nu}\partial_{\mu}\phi_i\partial_{\nu}\phi_i
-m^2\phi_i^2\right)
+ ~\frac{\kappa}{2}
\left(
\widetilde{h}^{\mu\nu}\partial_{\mu}\phi_i\partial_{\nu}\phi_i
- m^2\widetilde{h}\phi_i^2
\right)\nonumber\\
&&+ ~\frac{\kappa^2m^2}{4}
\left(
\widetilde{h}^{\mu\nu}\widetilde{h}_{\mu\nu}-\widetilde{h}^2
\right)
\phi_i^2
+{\rm O}(\kappa^3).
\end{eqnarray}
All the scalar one-loop diagrams (fig.2) have degrees of divergence
three. By using the Pauli-Villars-Gupta regularization, 
the counterterms are found to be \cite{Kugo}
\begin{eqnarray}
{\cal L}^{(0)}_{\rm{count}}&=&- ~\frac{1}{24\pi}
\left(
(2-\sqrt{2})\Lambda^3- ~\frac{3}{\sqrt{2}}\Lambda m^2+2m^3
\right)\sqrt{g}\nonumber\\
&&- ~\frac{1}{24\pi}\left(
(\sqrt{2}-1)\Lambda-m
\right)\sqrt{g}R \label{eq:count}
\end{eqnarray}
with a large regulator mass $\Lambda$. The coefficients 
in (\ref{eq:count}) are so determined that 
the renormalized graviton $(\widetilde{h}^{\mu\nu})$ two-point function 
$\Gamma^{(2)}_{\mu\rho,\nu\sigma}(p)$ 
should satisfy the following renormalization conditions:
\begin{eqnarray}
\left.\Gamma^{(2)}_{\mu\rho,\nu\sigma}(p)\right|_{p=0}&=&0,
\label{eq:Rcondition0-1}\\
\left. \frac{\partial\Gamma^{(2)}_{\mu\rho,\nu\sigma}
}{\partial p^2}(p)\right|_{p=0}
&=& \frac{1}{2}\left(
P^{(2)}_{\mu\rho,\nu\sigma}-Q_{\mu\rho,\nu\sigma}
\right), \label{eq:Rcondition0-2}
\end{eqnarray}
where $P^{(2)}_{\mu\rho,\nu\sigma}$ denotes 
the physical spin-2 projection
operator
\begin{equation}
P^{(2)}_{\mu\rho,\nu\sigma}
= \frac{1}{2}(\theta_{\mu\nu}\theta_{\rho\sigma}
+\theta_{\mu\sigma}\theta_{\nu\rho}-\theta_{\mu\rho}\theta_{\nu\sigma}),
\end{equation}
\begin{equation}
\theta_{\mu\nu}=\eta_{\mu\nu}- ~\frac{p_{\mu}p_{\nu}}{p^2},
\end{equation}
and the explicit form of the other remaining gauge-dependent parts 
$Q_{\mu\rho,\nu\sigma}$ is presented in appendix A. 
The right-hand side of (\ref{eq:Rcondition0-2}) is just a tree-level
contribution from $\sqrt{g}R$, 
and the local gauge symmetry is not yet
fixed.  
The general covariance ensures 
that the choice of counterterms (\ref{eq:count}) 
also makes other diagrams in fig.2 finite, and in particular the
tadpoles vanish.
Since there are no infrared divergences and the scalar mass does not 
affect the ultraviolet behavior of 
$\Gamma^{(2)}_{\mu\rho,\nu\sigma}(p)$,
we will henceforth set $m=0$ for simplicity. 
In this case the renormalized vacuum polarization tensor 
$\Pi_{\mu\rho,\nu\sigma}^{\rm{scalar}}$ is calculated to be 
 \cite{Kugo}
\begin{equation}
\Pi_{\mu\rho,\nu\sigma}^{\rm{scalar}}
=+ ~\frac{\kappa^2(-p^2)^{ \frac{3}{2}}}{512}
(P^{(2)}_{\mu\rho,\nu\sigma}+2Q_{\mu\rho,\nu\sigma}). 
\label{eq:Pi_scalar}
\end{equation}
The graviton propagator thus reads 
\begin{equation}
<\widetilde{h}_{\mu\rho}(p)\widetilde{h}_{\nu\sigma}(-p)>=i\left[
 \frac{p^2}{2}\left(
P^{(2)}_{\mu\rho,\nu\sigma}-Q_{\mu\rho,\nu\sigma}\right)
+ ~\frac{1}{2}\Pi_{\mu\rho,\nu\sigma}^{\rm{scalar}}
+(\mbox{\rm gauge fixing})
\right]^{-1}.
\end{equation}
The tensor structure of $\Pi_{\mu\rho,\nu\sigma}^{\rm{scalar}}$ 
is expressed in terms of $P^{(2)}_{\mu\rho,\nu\sigma}$ and 
$Q_{\mu\rho,\nu\sigma}$ only since the matter integration is performed 
in a gauge-invariant way.  If we take the harmonic gauge 
$\partial_{\mu}\widetilde{h}^{\mu\nu}=0$, we have 
\begin{equation}
<\widetilde{h}_{\mu\rho}(p)\widetilde{h}_{\nu\sigma}(-p)>=
i\left[
 \frac{2P^{(2)}_{\mu\rho,\nu\sigma}}{p^2+
                        ~\frac{\kappa^2}{512}(-p^2)^{\frac{3}{2}}}
+ ~\frac{2P^{(0-\rm{s})}_{\mu\rho,\nu\sigma}}{-p^2+
                         ~\frac{\kappa^2}{256}(-p^2)^{\frac{3}{2}}}
\right].
\label{eq:propagator}
\end{equation}
The first term in (\ref{eq:propagator}) is gauge-independent, while 
the second term 
depends on the gauge choice. 
The definition of $P^{(0-\rm{s})}_{\mu\rho,\nu\sigma}$ is given in 
appendix A. 
 
        In view of (\ref{eq:propagator}), the graviton propagator
behaves like $p_{\rm{E}}^{-3}$ $(p_{\rm{E}}\equiv\sqrt{-p^2})$ for a 
large momentum. Also, 
all $n$-point vertices ($n\geq 3$) scale like $p_{\rm{E}}^{-3}$ 
in the ultraviolet
regime, as expected from Weinberg's theorem
 \cite{Weinbergtheorem}.
One may then easily see that this
expansion is power-counting renormalizable. Furthermore, the standard
argument in gauge theories may formally 
apply to prove the all-order renormalizability  \cite{Kugo}. 
Namely, one may first derive the Ward-Takahashi (Slavnov-Taylor)
identity as a consequence of the BRS invariance. It leads to a 
renormalization equation, which governs the possible structure of the
divergent part of the proper vertices. Then, by induction, it is 
shown
that the solution of the equation 
is a BRS-invariant local functional 
of dimension three or less. In this way the $1/N$ expansion 
is claimed to be renormalizable  \cite{Kugo}. 

\subsection{A spin-2 tachyonic pole in the graviton propagator}

\indent

        The above proof of renormalizability 
is formal and does not actually work, 
since the graviton propagator (\ref{eq:propagator})
develops a spin-2 tachyonic pole in $p_{\rm{E}}$ at  
$p_{\rm{E}}=512\kappa^{-2}$ which renders the higher-order  
calculations ill-defined. 

        The presence of a tachyonic pole at the Planck scale may be
seen as a reflection of the breakdown of the $1/N$ approximation
based on (\ref{eq:starting}) at that scale. 
To see this let us calculate the renormalization-group
$\beta$ function. 
We define the $\beta$ function by setting the following
renormalization condition:
\begin{equation}
\left. \frac{\partial\Gamma^{(2)}_{\mu\rho,\nu\sigma}
}{\partial p^2}(p)\right|_{-p^2=\mu^2}
= \frac{1}{2}\left(
P^{(2)}_{\mu\rho,\nu\sigma}+\cdots
\right), \label{eq:Rcondition1-2}
\end{equation}
by which we replace the condition (\ref{eq:Rcondition0-2}). 
Here $\mu$
is an arbitrary renormalization mass scale, and the ellipsis denotes 
the terms proportional to $Q_{\mu\rho,\nu\sigma}$. 
In this case the counterterms are given by  
\begin{eqnarray}
{\cal L}^{(0)}_{\rm{count}}&=&- ~\frac{1}{24\pi}
(2-\sqrt{2})\Lambda^3\sqrt{g}\nonumber\\
&&+\left\{- ~\frac{1}{24\pi}
(\sqrt{2}-1)\Lambda
+ ~\frac{3\mu}{1024}\right\}
\sqrt{g}R. \label{eq:count'}
\end{eqnarray}
The $\beta$ function is obtained from the equations 
\begin{equation}
\mu\frac{\partial}{\partial\mu}\kappa_0^2=0,
\end{equation}
\begin{equation}
\frac{1}{\kappa_0^2}= \frac{\mu}{\kappa^2}
+\left\{- ~\frac{1}{24\pi}
(\sqrt{2}-1)\Lambda
+ ~\frac{3\mu}{1024}\right\},
\end{equation}
where $\kappa_0$ is the bare coupling constant.
They provide 
\begin{eqnarray}
\beta(\kappa^2)&\equiv&
\mu\frac{\partial}{\partial\mu}\kappa^2\nonumber\\
&=&\kappa^2+ ~\frac{3}{1024}\kappa^4.\label{eq:betafn}
\end{eqnarray}
This indicates that the theory is asymptotically
non-free and has no ultraviolet fixed points in the large $N$ limit. 
By integrating the leading-order $\beta$ function (\ref{eq:betafn}), 
we obtain the following running coupling constant $\kappa^2(\mu)$:
\begin{equation}
\frac{\kappa^2(\mu)}{\kappa^2(\mu)
+ \frac{1024}{3}}\raisebox{-2mm}{\Huge /}
 \frac{\kappa^2(\mu_0)}{\kappa^2(\mu_0)+ \frac{1024}{3}}
= \frac{\mu}{\mu_0}
\end{equation}
or
\begin{equation}
\kappa^2(\mu)
= \frac{\frac{1024}{3}}{A\mu^{-1}-1},\label{eq:running}
\end{equation}
where $\mu_0$ is an integration constant, and 
$A= \frac{\kappa^2(\mu_0)+ \frac{1024}{3}}{\kappa^2(\mu_0)}\mu_0$. 
We see from (\ref{eq:running}) that $\kappa^2(\mu)$ diverges
if $\mu$ approaches to some finite value $A$.  Hence   
the expansion in terms of the dressed propagator 
(\ref{eq:propagator})
is not reliable in the ultraviolet regime. 
Therefore it is plausible to consider that 
the appearance of the tachyonic pole in (\ref{eq:propagator}) is 
a reflection of the breakdown of this expansion 
in the present system. 
The situation is similar to those in other well-known 
non-asymptotically-free theories without fixed points, such as  
four-dimensional QED  and $\phi^4$ theories \cite{phi4}.

        To understand the situation it would be worth comparing it  
with three-  \hskip -2mm dimensional QCD (QCD${}_3$). QCD${}_3$ is a
super-renormalizable asymptotically free theory, and much reliance
cannot be placed on the loop approximation in the {\em low}-momentum
region. 
In this case the dressed ghost and gluon propagators also 
exhibit tachyonic poles in the {\em infrared} regime \cite{AP}, 
which implies the breakdown of the validity 
of the dressed propagators. 
In this sense the $\mu_0$-dependent constant $A$ above may be
understood as an analogue of $\Lambda_{\rm{QCD}}$ parameter.

\section{Inclusion of other unitary matter fields}
\indent

In sect.II we have seen that the $1/N$ expansion of
three-dimensional gravity coupled to $N$ scalar fields is
unstable due to the appearance of the tachyonic mode in the graviton
propagator. In this section we consider the cases of other unitary
matter fields, searching for a stable and renormalizable expansion 
of three-dimensional gravity. 
The Feynman rules for the calculations in this section are
listed in Table.1.

\subsection{Massless spinors}

\indent

        Let us first consider the coupling with massless spinors. 
The matter lagrangian is given by 

\begin{equation}
{\cal L}_{\rm{spinor}}= \frac{e}{2}ie_a^{~\mu}
\left( 
\overline{\psi}\gamma^a\partial_{\mu}\psi
-\partial_{\mu}\overline{\psi}\gamma^a\psi
+ ~\frac{i}{2}\overline{\psi}\omega_{\mu bc}\epsilon^{abc}\psi
\right), \label{eq:Lfermion}
\end{equation}
where $\psi$ is an $N$-plet spinor and we omit their indices for a
simple notation. The dreibein $e^a_{~\mu}$ and the spin-connection
$\omega_{\mu}^{~ab}$ are defined by 
\begin{equation}
g_{\mu\nu}=e^a_{~\mu}e_{a\nu}
=e^a_{~\mu}e^b_{~\nu}\eta_{ab},~~ e=\det e^a_{~\mu}
\end{equation}
and
\begin{equation}
\omega_{\mu}^{~ab}= \frac{1}{2}e^{a\nu}
(\partial_{\mu} e^b_{~\nu}-\partial_{\nu}e^b_{~\mu})
+ ~\frac{1}{4}e^{a\nu}e^{b\lambda}
(\partial_{\lambda}e^c_{~\nu}-\partial_{\nu}e^c_{~\lambda})e_{c\mu}
-(a\leftrightarrow b).
\end{equation}
Replacing ${\cal L}_{\rm{matter}}$ in (\ref{eq:totalL}) by
(\ref{eq:Lfermion}) and integrating over $\psi$ fields, 
we have the spinor one-loop effective
action 
\begin{eqnarray}
\Gamma
&=&\hskip -2mm N\left[\int d^3x\left( \frac{1}{\kappa^2}\sqrt{g}R+
\frac{e}{2}ie_a^{~\mu}
\left( 
\overline{\psi}\gamma^a\partial_{\mu}\psi
-\partial_{\mu}\overline{\psi}\gamma^a\psi
+\frac{i}{2}\overline{\psi}\omega_{\mu bc}\epsilon^{abc}\psi
\right)+
{\cal L}_{\rm{FP+GF}}\right)\right.\nonumber\\
&&\left.-i\log{\rm Det}(-D(g_{\mu\nu}))+\int d^3x
{\cal L}^{(0)}_{\rm{count}}\right],
\end{eqnarray}
where we have made appropriate rescaling of $\kappa$, $\psi$, 
ghosts and gauge parameters as in sect.II. 
$D(g_{\mu\nu})$ denotes the kernel of the spinor bilinear
in ${\cal L}_{\rm{spinor}}$. 

        To obtain the functional determinant we expand 
the dreibein as  
\begin{equation}
e^a_{~\mu}=\delta^a_{~\mu}+\kappa f^a_{~\mu} .
\end{equation}
In terms of $f^a_{~\mu}$, ${\cal L}_{\rm{spinor}}$ is expanded as 
follows:
\begin{eqnarray}
{\cal L}_{\rm{spinor}}&=& \frac{i}{2}
(\overline{\psi}\gamma^a\partial_a\psi
-\partial_a\overline{\psi}\gamma^a\psi)\nonumber\\
&&+ ~\frac{i}{2}\kappa(-f_a^{~\mu}+f\delta_a^{~\mu})
(\overline{\psi}\gamma^a\partial_a\psi
-\partial_a\overline{\psi}\gamma^a\psi)\nonumber\\
&&+{\rm O}(\kappa^2)
+\mbox{\rm (functionals of the antisymmetric part of $f^a_{~\mu}$)}
\end{eqnarray}
with $f=f^a_{~a}$. It suffices to consider only the symmetric part of
$f^a_{~\mu}$, which is related to $h_{\mu\nu}$ by
\begin{equation} 
f^a_{~\mu}= \frac{1}{2}h^a_{~\mu}.\label{eq:f->h}
\end{equation}
For actual calculations it is more convenient to work in terms of the
linear combination 
\begin{equation}
\widetilde{f}_a^{~\mu}\equiv -f_a^{~\mu}+f\delta_a^{~\mu}.
\label{eq:tf}
\end{equation}
We will calculate the spinor
one-loop diagram (fig.3) by the dimensional regularization. 
By this regularization we obtain a finite result without any
subtractions because of the odd-dimensionality. Since the 
spinors are massless, we have no dimensionful constants except for
$\kappa$, and so the renormalization conditions
(\ref{eq:Rcondition0-1}) and (\ref{eq:Rcondition0-2}) are 
automatically
satisfied. 
It is known that in three dimensions massive spinors 
coupled to gravity induces gravitational Chern-Simons terms if the 
dimensional regularization is used \cite{gravitationalCS1}.  
One would also obtain those parity-odd terms 
if one employed the Pauli-Villars regularization for massless
spinors, and the graviton becomes massive
topologically \cite{gravitationalCS2}.  
In this paper we will make use of the dimensional regularization for
massless spinors so that no Chern-Simons terms are induced, and will
not consider the effect of those terms.

A straightforward spinor one-loop calculation provides the following 
graviton-graviton vacuum polarization tensor:
\footnote{The expressions of 
$\Pi_{\mu\rho,\nu\sigma}^{\rm{scalar}}$ and 
$\Pi_{\mu\rho,\nu\sigma}^{\rm{spinor}}$  
may be found in refs.\cite{Pi1,Pi2}.}
\begin{equation}
\Pi_{\mu\rho,\nu\sigma}^{\rm{spinor}}
=+ ~\frac{\kappa^2(-p^2)^{\frac{3}{2}}}
{256}P^{(2)}_{\mu\rho,\nu\sigma},
\label{eq:Pi_spinor}
\end{equation}
where, by using the formulas in appendix B, 
the two external graviton fields of (\ref{eq:Pi_spinor}) have been  
transformed from $\widetilde{f}^a_{~\mu}$ to $\widetilde{h}^{\mu\nu}$ 
so that we may compare this with (\ref{eq:Pi_scalar}). 
Note that in this case are induced no terms proportional to
$Q_{\mu\rho,\nu\sigma}$.   
        We see that 
the sign of $(-p^2)^{\frac{3}{2}}P^{(2)}_{\mu\rho,\nu\sigma}$ 
is plus, meaning the presence of a spin-2 tachyon.\footnote
{
We were informed by T. Kugo that he had also obtained this
fact \cite{Kugo2}.  
}

\subsection{U(1) gauge fields}

\indent

        We next consider the coupling to the 
$({\rm U}(1))^N$ gauge fields, {\it i.e.} $N$ independent abelian 
gauge fields. The matter lagrangians are the following: 
 \begin{eqnarray}
{\cal L}_{\rm{U(1)}}
&=&- ~\frac{1}{4}\sqrt{g}g^{\mu\nu}g^{\lambda\sigma}F_{\mu\lambda}F_{\nu\sigma}
+{\cal L}_{\rm{U(1)GF}}, 
\label{eq:L_U(1)}\\
{\cal L}_{\rm{U(1)GF}}
&=&- ~\frac{1}{2\alpha}\sqrt{g}(\nabla^{\mu}A_{\mu})^2,
\label{eq:L_U(1)GF}\\
{\cal L}_{\rm{U(1)FP}}
&=&\sqrt{g}i\overline{c}\nabla^{\mu}\partial_{\mu}c,
\label{eq:L_U(1)FP}
\end{eqnarray}
where $A_{\mu}=A_{\mu}^{(i)}$, $c=c^{(i)}$ and 
$\overline{c}=\overline{c}^{(i)}$ $(i=1,\ldots,N)$.  
$\alpha$ is a gauge parameter.  
If the space-time is flat,
U(1) FP ghosts are free and completely 
decoupled. In our case, however, the ghosts do interact with
gravitational field and can not be neglected. 

        Repeating a similar procedure, we have only to calculate two
diagrams shown in fig.4 to examine the pole structure of the dressed
graviton propagator. In this case we expand 
${\cal L}_{\rm{U(1)}}$ in terms
of $h_{\mu\nu}$:
\begin{eqnarray}
{\cal L}_{\rm{U(1)}}
&=& \frac{1}{2}A^{\lambda}
(\eta_{\mu\lambda}\raisebox{-0.5mm}{\hskip 0.6mm\rule{0.1mm}{3.1mm}\rule{3.1mm}
{0.1mm}\rule{0.1mm}{3.2mm}\hskip
-3.3mm\rule[3.1mm]{3.2mm}{0.1mm}\hskip 0.7mm}-\partial_{\mu}\partial_{\lambda})
A^{\mu}
- ~\frac{1}{2\alpha}(\partial_{\mu}A^{\mu})^2\nonumber\\
&&- ~\frac{\kappa}{2}h_{\alpha\beta}
\left[
~\frac{1}{2}\eta^{\alpha\beta}\partial_{\mu}A_{\lambda}
(\eta^{\lambda\sigma}\partial^{\mu}-\eta^{\mu\sigma}\partial^{\lambda})A_{\sigma}
\right.\nonumber\\
&&-\partial^{\alpha}A_{\lambda}
(\eta^{\lambda\sigma}\partial^{\beta}-\eta^{\beta\sigma}
\partial^{\lambda})A_{\sigma}\nonumber\\
&&\left.
+\partial_{\mu}A_{\lambda}\eta^{\lambda\alpha}
(\eta^{\mu\sigma}\partial^{\beta}-\eta^{\beta\sigma}\partial^{\mu})A_{\sigma}
\phantom{\frac{1}{2}}\!\!\!\right]\nonumber\\
&&- ~\frac{\kappa}{2} \frac{1}{2\alpha}
\left(h\partial^{\lambda}A_{\lambda}\partial^{\sigma}A_{\sigma}
+2\partial^{\lambda}h\cdot A_{\lambda}\partial^{\sigma}A_{\sigma}\right)\nonumber\\
&&+{\rm O}(\kappa^2).
\end{eqnarray}
The most convenient gauge-choice is the Feynman gauge
$\alpha=1$. After some straightforward calculations we obtain 
\begin{equation}
\Pi_{\mu\rho,\nu\sigma}^{\rm{U(1)gauge}}
=+ ~\frac{3\kappa^2(-p^2)^{\frac{3}{2}}}{512}
(P^{(2)}_{\mu\rho,\nu\sigma}+2Q_{\mu\rho,\nu\sigma}).\label{eq:Pi_U1}
\end{equation} 
Again, this expression is that for the external $\widetilde{h}^{\mu\nu}$ 
fields.
As for the ghosts, ${\cal L}_{\rm{U(1)FP}}$ is conveniently
expanded by $\widetilde{h}^{\mu\nu}$:
\begin{equation}
{\cal L}_{\rm{U(1)FP}}=i\overline{c}\raisebox{-0.5mm}{\hskip 0.6mm\rule{0.1mm}{3.1mm}\rule{3.1mm}
{0.1mm}\rule{0.1mm}{3.2mm}\hskip
-3.3mm\rule[3.1mm]{3.2mm}{0.1mm}\hskip 0.7mm} c
-\kappa\widetilde{h}^{\mu\rho}i\partial_{\mu}\overline{c}\partial_{\rho}c
+{\rm O}(\kappa^2).
\end{equation}
A similar calculation gives 
\begin{equation}
\Pi_{\mu\rho,\nu\sigma}^{\rm{U(1)FP}}
=- ~\frac{2\kappa^2(-p^2)^{\frac{3}{2}}}{512}
(P^{(2)}_{\mu\rho,\nu\sigma}+2Q_{\mu\rho,\nu\sigma}).\label{eq:Pi_U1FP}
\end{equation}
Combining (\ref{eq:Pi_U1}) and (\ref{eq:Pi_U1FP}), we have the total
vacuum polarization tensor 
\begin{equation}
\Pi_{\mu\rho,\nu\sigma}^{\rm{U(1)total}}=
+ ~\frac{\kappa^2(-p^2)^{\frac{3}{2}}}{512}
(P^{(2)}_{\mu\rho,\nu\sigma}+2Q_{\mu\rho,\nu\sigma}). 
\end{equation} 
We find that the spin-2 part is the
same as that of the scalar case (\ref{eq:Pi_scalar}).

\subsection{Non-minimal scalar fields}

\indent

        Finally, let us consider $N$ scalar fields with a non-minimal
coupling to gravity. The lagrangian is given by 
\begin{eqnarray}
{\cal L}_{\rm{non-minimal}}
&=& \frac{1}{2}\sqrt{g}g^{\mu\nu}\partial_{\mu}\phi 
\partial_{\nu}\phi
+\lambda\sqrt{g}R\phi^2
\nonumber\\
&=&
- ~\frac{1}{2}\phi\raisebox{-0.5mm}{\hskip 0.6mm\rule{0.1mm}{3.1mm}\rule{3.1mm}
{0.1mm}\rule{0.1mm}{3.2mm}\hskip
-3.3mm\rule[3.1mm]{3.2mm}{0.1mm}\hskip 0.7mm}\phi
+\kappa\left[
\frac{1}{2}\widetilde{h}^{\mu\nu}\partial_{\mu}\phi\partial_{\nu}\phi
+\lambda\left(
\raisebox{-0.5mm}{\hskip 0.6mm\rule{0.1mm}{3.1mm}\rule{3.1mm}
{0.1mm}\rule{0.1mm}{3.2mm}\hskip
-3.3mm\rule[3.1mm]{3.2mm}{0.1mm}\hskip 0.7mm}\widetilde{h}+\partial_{\mu}\partial_{\nu}\widetilde{h}^{\mu\nu}
\right)\phi^2
\right]\nonumber\\
&&+{\rm O}(\kappa^2),
\end{eqnarray}
where $\lambda$ is a free real parameter. In this case the vacuum
polarization tensor is modified from (\ref{eq:Pi_scalar}) 
as follows: 
\begin{equation}
\Pi_{\mu\rho,\nu\sigma}^{\rm{non-minimal}}
=+ ~\frac{\kappa^2(-p^2)^{\frac{3}{2}}}{512}
\left(P^{(2)}_{\mu\rho,\nu\sigma}+
2(1+16\lambda)^2
Q_{\mu\rho,\nu\sigma}\right). 
\label{eq:Pi_nonminimal}
\end{equation}
Comparing (\ref{eq:Pi_nonminimal}) with (\ref{eq:Pi_scalar}), 
we see that the difference is only in the $Q_{\mu\rho,\nu\sigma}$ part. 
Therefore the inclusion of a non-minimal coupling can not cure the
spin-2 tachyon disease.
However, we will discuss in subsect.V.B its special role 
in avoiding the 
tachyon in the gauge-{\em dependent} (spin-0 or -1) part of the 
graviton propagator.

\subsection{Remarks}

\indent

        We have seen that the $1/N$ expansion of
three-dimensional Einstein gravity coupled to realistic matters 
such as (non-minimal) scalars, spinors and U(1) gauge fields, is
unstable; in all cases above the dressed graviton propagators 
possess tachyonic poles in the spin-2 part. 
The vacuum polarization tensor for 
gravitons is proportional to $(-p^2)^{\frac{3}{2}}$ and the
coefficient of $P^{(2)}_{\mu\rho,\nu\sigma}$ is
quantized to $\kappa^2(-p^2)^{\frac{3}{2}}/512$
times the physical degrees of freedom. 
For instance, in the case of U(1) gauge field the contributions to
$\Pi_{\mu\rho,\nu\sigma}^{\rm{U(1)total}}$ of 
photon itself is three in this unit, but the FP ghosts cancel two
unphysical contributions from longitudinal and scalar modes, 
and that leaves precisely one: the physical degrees of freedom 
of photon in three dimensions. 
On the other hand, as we have seen in the previous section, 
a tachyonic pole in the dressed propagator
necessarily follows from such positive contributions in this unit. 

Concluding sects. II and III, in the $1/N$ expansion of
three-dimensional Einstein gravity coupled 
to $N$-plet {\em unitary} matters
(the matters with positive degrees of freedom) 
a tachyonic pole exists
in the spin-2 mode of the leading graviton propagator.
Physically, this implies that the flat space-time 
$g_{\mu\nu}=\eta_{\mu\nu}$, having been taken as the classical
vacuum of space-time (\ref{eq:flat}), is quantum-mechanically 
unstable if the number $N$ of unitary matters is very large.
Theoretically, the presence of a tachyon prevents us to 
include graviton-loop corrections consistently.

As mentioned in subsect.II.B, the presence of a tachyon in the 
spin-2 part of the graviton propagator has a close correspondence 
with the absence of an ultraviolet fixed point for $\kappa$. 
Note here that in the ($2+\epsilon$)-dimensional gravity \cite{KKN} 
the one-loop $\beta$ function ($16\pi G=\kappa^2$)
\begin{equation}
\beta(G)=\epsilon G- \frac{25-c}{24\pi}G^2,
\end{equation}  
exhibits an ultraviolet fixed point only when the central charge 
$c$ of matters satisfies $c<25$. 
This fact may suggest that {\em in three dimensions also, the
existence of the non-trivial ultraviolet fixed point provides 
 an upper limit for the number of physical degrees of freedom
for matters.}
This conjecture is at least consistent with our results that 
three-dimensional gravity coupled to infinite number of unitary
matters has no ultraviolet fixed point.
Thus, {\em if we adhere to Einstein gravity} $\sqrt{g}R$
and want to stabilize the flat space-time in the present $1/N$ 
scheme, it would be necessary to consider in some sense a coupling 
to matters with {\em negative} degrees of freedom; otherwise 
we should modify the gravity part of the lagrangian.

\section{Removing tachyon in the spin-2 part}

\indent

Based on the above results,  we will discuss in sects. IV and V
what could be the possible modification or extension
of models to stabilize the flat
space-time, i.e. to remove tachyons in the graviton propagator. 
In this section we indicate two possibilities to circumvent the
presence of a tachyon in the {\em spin-2 part} of the propagator.

\subsection{Higher-derivative term for gravity}

\indent

The first possibility is the generalization of gravity 
by adding higher-derivative terms such as $\beta R^2$ 
with matter contents unchanged. In the history of quantum gravity, 
higher-derivative gravity has sometimes been rejected by the 
reason that it may not keep unitarity. However, in the broad viewpoint 
that we may look at gravity as a certain low-energy effective 
field theory, this point need not be seriously taken.
The unitarity problem should be addressed in an original complete
theory such as superstrings, valid in all energy scales. 
In the low-energy renormalizable theory, however, we can still
legitimately argue infrared properties such as a phase structure of 
space-time. 

Due to this generalization the pole structure of a dressed graviton
propagator shall be changed and a possibility arises to have 
a tachyon-free propagator by choosing a suitable value for a
new parameter $\beta$. 
This generalization may indeed be a most natural direction
for treating gravity coupled to {\em unitary} matters.
The graviton fluctuations in higher orders can be consistently 
incorporated to search for an ultraviolet fixed point.
Although its absence to leading order remains, 
it is very interesting to see to which direction the graviton-loop 
effect works in the renormalization of $\kappa$. 

Although this approach may in fact lead to a tachyon-free theory, 
we shall instead lose the theoretically fascinating possibility that 
quantum gravity in three dimensions could be controllable by a {\em
single} parameter $\kappa$, namely that Einstein gravity itself may
be renormalizable.
In this paper we mainly stick to Einstein gravity $\sqrt{g}R$, 
and will not pursue the higher-derivative theory which we would
like to consider in a future.

\subsection{Non-unitary matters}
\label{subsect:Nonunitarymatters}

\indent

The second possibility to avoid a spin-2 tachyon is to couple 
Einstein gravity to {\em non-unitary matters}.
If the conjecture presented in subsect.III.D is correct, 
an ultraviolet fixed point should exist in the 
$N_{\rm matter}\rightarrow -\infty$ limit.
This limit is analogous to the $c\rightarrow -\infty$ limit in the 
two- and $(2+\epsilon)$-dimensional gravity. 
In two dimensions the $c\rightarrow-\infty$ limit is indeed known as
the semi-classical limit of gravity \cite{DNOPZ}
and its knowledge has been useful when the correct exact solution is
chosen from two branches \cite{KPZDDK}
\footnote{As one of the other interesting examples 
of two-dimensional 
gravity coupled to non-unitary matters, a numerical simulation has 
recently been performed for $c=-2$ and the 
fractal scaling has been clearly observed \cite{KKSW}.}.
In our three-dimensional model this formal limit can be realized by
coupling gravity to $N$-plet ghost matters and taking the 
$N\rightarrow\infty$ limit.
The physical degrees of freedom of matters are negative, 
and from our preceding analyses follows that the tachyonic 
pole in the spin-2 part of the graviton propagator should disappear.
Although the positivity of the Hilbert space is lost, 
this model may actually serve as an interesting theoretical model that
could also be simulated numerically.  

As an explicit example, consider the minimal coupling of gravity 
to massless fermionic scalar fields, described by the lagrangian
\begin{equation}
{\cal L}_{\rm matter}=\sqrt{g}g^{\mu\nu}\partial_{\mu}\bar{c}_i
\partial_{\nu}c_i,
\label{eq:NUmatter}
\end{equation}
where $c_i$ and $\bar{c}_i$ are the $N$-plet anti-commuting
scalar and their conjugate fields. Integrating over these 
non-unitary (ghost) matters one can formulate the $1/N$ expansion 
of the
system ${\cal L}$ in eq.(\ref{eq:totalL}) where
${\cal L}_{\rm matter}$ is replaced by eq.(\ref{eq:NUmatter}).
If we use the dimensional regularization and normalize the leading 
graviton two-point function at $p^2=0$ by 
eqs.(\ref{eq:Rcondition0-1})
and (\ref{eq:Rcondition0-2}), 
the cosmological constant is renormalized to zero and we get
($\kappa_0^2=\kappa^2\mu^{-1}$)
\begin{equation}
\Gamma^{(2)}_{\mu\rho,\nu\sigma}({p})=\frac{p^2}{2}\left[
1+\frac{\kappa^2\mu^{-1}}{512}\left(-p^2\right)^{1\over2}\right]
P^{(2)}_{\mu\rho,\nu\sigma}+\cdots.
\label{eq:renormalized2point1}
\end{equation}
Due to the square root term $(-p^2)^{1/2}$, $p^2=0$ is a branch point
and $\Gamma^{(2)}_{\mu\rho,\nu\sigma}({p})$ is 
a double-valued function on
the complex $p^2$ plane made up of two Riemann sheets. 
Apart from the expected zero at $p^2=0$, no other zeros are
in the first sheet. No tachyonic poles then
exist in the spin-2 part of the propagator.
If we instead normalize $\Gamma^{(2)}_{\mu\rho,\nu\sigma}({p})$ by
introducing a finite renormalization mass scale $\mu$ like in
(\ref{eq:Rcondition1-2}), we shall obtain the renormalization-group
$\beta$ function exhibiting the non-trivial ultraviolet fixed
point for $\kappa^2(\mu)$\footnote{More correctly speaking, a fixed
point should exist for the original coupling constant 
(=$\kappa^2N$ for the present $\kappa^2$), as is usual
for standard analyses using $1/N$ expansion.},
 which would suggest the existence of two 
gravitational phases.
This issue seems very interesting, but includes a non-trivial 
subtlety 
with respect to the metric $g_{\mu\nu}$  redefinition ambiguity. 
The detailed analyses will be given elsewhere \cite{MY2}.

\section{Removing tachyon in the
gauge-dependent \newline (spin-1 or -0) part}

\indent

As we have argued above, even if we keep Einstein gravity
$\sqrt{g}R$, there exists a case where a tachyon can be avoided 
in the spin-2 part, i.e. the gauge-invariant part of the graviton 
propagator. Although the gauge-dependent piece of the propagator 
indicated by the ellipsis above, i.e. the part dependent on the 
choice of gauge fixing, can not affect any gauge-invariant 
quantity, the presence of a tachyonic pole in the gauge part also 
spoils the calculability to higher orders.
It is therefore important to study whether one can avoid tachyonic
poles in the gauge-dependent part as well.  
In this section we first investigate explicitly the pole 
structure of the gauge-dependent part of the propagator in the 
second possibility (subsect.\ref{subsect:Nonunitarymatters}).
From all the results we will finally indicate the models which are
expected to be completely free from tachyons.

\subsection{Tachyon in the gauge-dependent part} 

\indent

On the assumption that we have chosen the second case 
(subsect.\ref{subsect:Nonunitarymatters}) 
so that a tachyon does not exist in the spin-2 part of the graviton 
propagator, we will examine here whether we may take any
gauge-fixing such that the graviton propagator would not possess any
spin-0 or -1 tachyonic poles, either. 
For this purpose, let us explicitly write the FP ghost and the
gauge-fixing lagrangians for the general coordinate gauge invariance 
\begin{eqnarray}
{\cal
L}_{\rm{FP}}
&=&i\overline{C}_{\mu}\partial_{\nu}D^{\mu\nu}_{~~\rho}C^{\rho},\\
{\cal
L}_{\rm{GF}}
&=&
- ~\frac{1}{2\xi}
g_{\mu\nu}
\widehat{e}(\raisebox{-0.5mm}{\hskip 0.6mm\rule{0.1mm}{3.1mm}\rule{3.1mm}
{0.1mm}\rule{0.1mm}{3.2mm}\hskip
-3.3mm\rule[3.1mm]{3.2mm}{0.1mm}\hskip 0.7mm})
\left(
\partial_{\lambda}\widetilde{g}^{\mu\lambda}
-\zeta\partial^{\mu}\sqrt{g}
\right)\cdot
\widehat{e}(\raisebox{-0.5mm}{\hskip 0.6mm\rule{0.1mm}{3.1mm}\rule{3.1mm}
{0.1mm}\rule{0.1mm}{3.2mm}\hskip
-3.3mm\rule[3.1mm]{3.2mm}{0.1mm}\hskip 0.7mm})
\left(
\partial_{\rho}\widetilde{g}^{\nu\rho}
-\zeta\partial^{\nu}\sqrt{g}
\right),
\label{eq:L_GF}
\\
D^{\mu\nu}_{~~\rho}
&=&
\widetilde{g}^{\mu\sigma}\delta_{\rho}^{\nu}\partial_{\sigma}
+\widetilde{g}^{\nu\sigma}\delta_{\rho}^{\mu}\partial_{\sigma}
-\widetilde{g}^{\mu\nu}\partial_{\rho}
-(\partial_{\rho}\widetilde{g}^{\mu\nu}),
\end{eqnarray}
where we have adopted the following linear gauge-fixing condition: 
\begin{equation}
\widehat{e}(\raisebox{-0.5mm}{\hskip 0.6mm\rule{0.1mm}{3.1mm}\rule{3.1mm}
{0.1mm}\rule{0.1mm}{3.2mm}\hskip
-3.3mm\rule[3.1mm]{3.2mm}{0.1mm}\hskip 0.7mm})
(
\partial_{\nu}\widetilde{g}^{\mu\nu}
-\zeta\partial^{\mu}\sqrt{g}
)
+ ~\frac{1}{2}\xi B^{\mu}=0
\end{equation}
with some function $\widehat{e}(\raisebox{-0.5mm}{\hskip 0.6mm\rule{0.1mm}{3.1mm}\rule{3.1mm}
{0.1mm}\rule{0.1mm}{3.2mm}\hskip
-3.3mm\rule[3.1mm]{3.2mm}{0.1mm}\hskip 0.7mm})$ of $\raisebox{-0.5mm}{\hskip 0.6mm\rule{0.1mm}{3.1mm}\rule{3.1mm}
{0.1mm}\rule{0.1mm}{3.2mm}\hskip
-3.3mm\rule[3.1mm]{3.2mm}{0.1mm}\hskip 0.7mm}$ and an auxiliary
field $B^{\mu}$. If one takes $\widehat{e}(\raisebox{-0.5mm}{\hskip 0.6mm\rule{0.1mm}{3.1mm}\rule{3.1mm}
{0.1mm}\rule{0.1mm}{3.2mm}\hskip
-3.3mm\rule[3.1mm]{3.2mm}{0.1mm}\hskip 0.7mm})=\mbox{\rm const.}$,
$\zeta=0$ and $\xi\rightarrow 0$, 
the gauge is reduced to the familiar de Donder
gauge. We will take $\widehat{e}(\raisebox{-0.5mm}{\hskip 0.6mm\rule{0.1mm}{3.1mm}\rule{3.1mm}
{0.1mm}\rule{0.1mm}{3.2mm}\hskip
-3.3mm\rule[3.1mm]{3.2mm}{0.1mm}\hskip 0.7mm})$ to be a non-constant 
function 
in order to improve the high-momentum behavior of the
gauge-dependent part of the propagator. For this purpose it suffices
to choose $\widehat{e}(\raisebox{-0.5mm}{\hskip 0.6mm\rule{0.1mm}{3.1mm}\rule{3.1mm}
{0.1mm}\rule{0.1mm}{3.2mm}\hskip
-3.3mm\rule[3.1mm]{3.2mm}{0.1mm}\hskip 0.7mm})$ to be a linear function of $\raisebox{-0.5mm}{\hskip 0.6mm\rule{0.1mm}{3.1mm}\rule{3.1mm}
{0.1mm}\rule{0.1mm}{3.2mm}\hskip
-3.3mm\rule[3.1mm]{3.2mm}{0.1mm}\hskip 0.7mm}$.
Note that the
operator $\widehat{e}(\raisebox{-0.5mm}{\hskip 0.6mm\rule{0.1mm}{3.1mm}\rule{3.1mm}
{0.1mm}\rule{0.1mm}{3.2mm}\hskip
-3.3mm\rule[3.1mm]{3.2mm}{0.1mm}\hskip 0.7mm})$ in the gauge-fixing does not affect the
kinetic part of the FP ghosts  
since the jacobian of the field redefinition 
$\raisebox{-0.5mm}{\hskip 0.6mm\rule{0.1mm}{3.1mm}\rule{3.1mm}
{0.1mm}\rule{0.1mm}{3.2mm}\hskip
-3.3mm\rule[3.1mm]{3.2mm}{0.1mm}\hskip 0.7mm}\overline{C}_{\mu}\rightarrow\overline{C}_{\mu}$ is
trivial \cite{AT}. 
        The effective action 
is invariant
under the BRS transformation 
\begin{eqnarray}
\delta_{\rm B}\widetilde{g}^{\mu\nu}&=&\kappa
D^{\mu\nu}_{~~\rho}C^{\rho},\nonumber\\ 
\delta_{\rm B}C^{\mu}&=&-\kappa C^{\nu}\partial_{\nu}C^{\mu},\nonumber\\
\delta_{\rm B}\overline{C}_{\mu}&=&iB_{\mu},\nonumber\\
\delta_{\rm B}B_{\mu}&=&0.
\end{eqnarray}
Indeed, up to irrelevant non-propagating terms,  
${\cal L}_{\rm{FP}}+{\cal L}_{\rm{GF}}$ 
is written in the following BRS-exact form \cite{KugoUehara}:
\begin{equation}
{\cal L}_{\rm{FP}}+{\cal L}_{\rm{GF}}
=-i\kappa^{-1}\delta_{\rm B}\left[
\overline{C}_{\mu}\left(
\widehat{e}(\raisebox{-0.5mm}{\hskip 0.6mm\rule{0.1mm}{3.1mm}\rule{3.1mm}
{0.1mm}\rule{0.1mm}{3.2mm}\hskip
-3.3mm\rule[3.1mm]{3.2mm}{0.1mm}\hskip 0.7mm})
(
\partial_{\nu}\widetilde{g}^{\mu\nu}
-\zeta\partial^{\mu}\sqrt{g}
)
+ ~\frac{1}{2}\xi B^{\mu}
\right)
\right].
\end{equation}

As understood from the results of previous sections, 
the quadratic parts of the Einstein action and 
of the one-loop effective action induced by the 
non-unitary matters specified by (\ref{eq:NUmatter}), reads 
in momentum space as 
\begin{eqnarray}
&&
\int d^3x \frac{1}{\kappa^2}\sqrt{g}R~~(\mbox{\rm quadratic in
$\widetilde{h}^{\mu\nu}$})\nonumber\\
&=&
\int \frac{d^3p}{(2\pi)^3} \frac{1}{2}
\widetilde{h}^{\mu\rho}(p)\widetilde{h}^{\nu\sigma}(-p)
\frac{p^2}{2}(P^{(2)}_{\mu\rho,\nu\sigma}-Q_{\mu\rho,\nu\sigma})
\label{eq:treeinmomspace}
\end{eqnarray}
and
\begin{eqnarray} 
&&
-i\log{\rm Det}(-D_{\rm{matter}}(g_{\mu\nu}))
+\int d^3x{\cal L}^{(0)}_{\rm{count}}
~~(\mbox{\rm quadratic in
$\widetilde{h}^{\mu\nu}$})
\nonumber\\
&=&
\int \frac{d^3p}{(2\pi)^3} \frac{1}{2}
\widetilde{h}^{\mu\rho}(p)\widetilde{h}^{\nu\sigma}(-p)
\left(-~\frac{\kappa^2(-p^2)^{\frac{3}{2}}}{512}\right)
(P^{(2)}_{\mu\rho,\nu\sigma}+2Q_{\mu\rho,\nu\sigma}),
\label{eq:1loopinmomspace}
\end{eqnarray}
respectively. The algebraic structure of the tensors in
the quadratic part in $\widetilde{h}^{\mu\nu}$ can conveniently be described by
$4\times 4$ matrices (See appendix A.). The coefficient of 
$\frac{1}{2}\widetilde{h}^{\mu\rho}(p)\widetilde{h}^{\nu\sigma}(-p)$
in the integrand of (\ref{eq:treeinmomspace}) can be written as 
\begin{equation}
\frac{p^2}{2}(P^{(2)}_{\mu\rho,\nu\sigma}-Q_{\mu\rho,\nu\sigma})=
\frac{p^2}{2}
\left[
\begin{array}{cccc}
1&&&\\
&0&&\\
&&-1&-\sqrt{2}\\
&&-\sqrt{2}&-2
\end{array}
\right],
\label{eq:tree}
\end{equation}
and that of (\ref{eq:1loopinmomspace}) reads 
\begin{equation}
- ~\frac{\kappa^2(-p^2)^{\frac{3}{2}}}{512}
(P^{(2)}_{\mu\rho,\nu\sigma}+2Q_{\mu\rho,\nu\sigma})
=
- ~\frac{\kappa^2(-p^2)^{\frac{3}{2}}}{512}
\left[
\begin{array}{cccc}
1&&&\\
&0&&\\
&&2&2\sqrt{2}\\
&&2\sqrt{2}&4
\end{array}
\right].
\label{eq:NUmatterone-loop}
\end{equation}
The addition of (\ref{eq:NUmatterone-loop}) to (\ref{eq:tree}) 
does not
change the rank of the matrix (\ref{eq:tree}), 
as it should be, due to the (general coordinate transformation) 
gauge-invariance of matter integration. 
Similarly, the quadratic part of ${\cal L}_{\rm{GF}}$ 
(\ref{eq:L_GF}) 
is written as 
\begin{eqnarray}
&&
\int d^3x
{\cal L}_{\rm{GF}}
~~(\mbox{\rm quadratic in
$\widetilde{h}^{\mu\nu}$})\nonumber\\
&=&
\int \frac{d^3p}{(2\pi)^3} \frac{1}{2}
\widetilde{h}^{\mu\rho}(p)\widetilde{h}^{\nu\sigma}(-p)
\cdot\left\{ - ~\frac{\kappa^2}{\xi}\right.\widehat{e}(-p^2)^2p^2\nonumber\\
&&\hskip -5mm\left.
\cdot
\left[
\frac{1}{2}P^{(1)}_{\mu\rho,\nu\sigma}
+2\zeta^2P^{\rm{(0-s)}}_{\mu\rho,\nu\sigma}
+\sqrt{2}(\zeta^2-\zeta)
(P^{\rm{(0-sw)}}_{\mu\rho,\nu\sigma}+P^{\rm{(0-ws)}}_{\mu\rho,\nu\sigma})
+(1-\zeta)^2P^{\rm{(0-w)}}_{\mu\rho,\nu\sigma}
\right]\right\}.\nonumber\\
&&
\end{eqnarray}
In the matrix notation $\{\cdots\}$ is represented by 
\begin{equation}
- ~\frac{\kappa^2}{\xi}\widehat{e}(-p^2)^2p^2
\left[
\begin{array}{cccc}     
0&&&\\
& \frac{1}{2}&&\\
&&2\zeta^2&\sqrt{2}(\zeta^2-\zeta)\\
&&\sqrt{2}(\zeta^2-\zeta)&(1-\zeta)^2
\end{array}
\right].
\label{eq:GF}
\end{equation}
The propagator is given by 
\begin{equation}
<\widetilde{h}_{\mu\rho}(p)\widetilde{h}_{\nu\sigma}(-p)>=i\left[
(\ref{eq:tree})+(\ref{eq:NUmatterone-loop})+(\ref{eq:GF})
\right]^{-1}.\label{eq:modprop}
\end{equation}
Hence no tachyonic pole appears in the spin-1 part if
$\widehat{e}(-p^2)$ 
has no zeroes in $p^2<0$. 
On the other hand, the spin-0 part of the propagator
(\ref{eq:modprop}) turns out to be the following form:
\begin{eqnarray}
&&
\left\{
\left(- ~\frac{p^2}{2}
- ~\frac{2\kappa^2(-p^2)^{\frac{3}{2}}}{512}
\right)
\left[
\begin{array}{cc}
1&\sqrt{2}\\
\sqrt{2}&2
\end{array}
\right]\right.
\nonumber\\
&&\left.
- ~\frac{\kappa^2}{\xi}\widehat{e}(-p^2)^2p^2
\left[
\begin{array}{cc}
2\zeta^2&\sqrt{2}(\zeta^2-\zeta)\\
\sqrt{2}(\zeta^2-\zeta)&(1-\zeta)^2
\end{array}
\right]\right\}^{-1}\nonumber\\
&=&
\frac{1}{(1+\zeta)^2}
\left\{
\left(- ~\frac{p^2}{2}
- ~\frac{2\kappa^2(-p^2)^{\frac{3}{2}}}{512}
\right)^{-1}
\left[
\begin{array}{cc}
(1-\zeta)^2&-\sqrt{2}(\zeta^2-\zeta)\\
-\sqrt{2}(\zeta^2-\zeta)&2\zeta^2
\end{array}
\right]\right.\nonumber\\
&&\left.
+\left(
- ~\frac{\kappa^2}{\xi}\widehat{e}(-p^2)^2p^2
\right)^{-1}
\left[
\begin{array}{cc}
2&-\sqrt{2}\\
-\sqrt{2}&1
\end{array}
\right]
\right\}.\label{eq:spin0}
\end{eqnarray}
In view of (\ref{eq:spin0}), the tachyonic pole appears at 
$p_{\rm{E}}=128\kappa^{-2}$. In fact, this pole can never be evaded 
by any choice of gauge parameters $(\xi, \zeta)$,
\footnote{
$\zeta=-1$ is not allowed since, if so, the spin-0 part of
(\ref{eq:GF}) becomes proportional to $Q_{\mu\rho,\nu\sigma}$, and hence
the gauge-fixing is incomplete.
}
or by any choice of
$\widehat{e}(-p^2)$. This is because each matrix 
of the first line of  
(\ref{eq:spin0}) has vanishing determinant, and consequently the
determinant of their sum always contains the factor 
$\left(-p^2/2
-2\kappa^2(-p^2)^{\frac{3}{2}}/512
\right)$.
Thus we conclude that we can never avoid a tachyonic pole in the
spin-0 or -1 part of the graviton propagator if we consider the 
$1/N$ expansion of three-dimensional gravity coupled to non-unitary 
matters only.

\subsection{Tachyon-free theories}

\indent 

Technically, the above tachyon observed in the gauge-dependent part
 comes from the fact that the coefficients of 
$P^{(2)}_{\mu\rho,\nu\sigma}$ and $Q_{\mu\rho,\nu\sigma}$ 
in the vacuum polarization tensor  have the
same signs, as is seen in (\ref{eq:NUmatterone-loop}). 
On the other hand, 
the vacuum polarization tensor 
$\Pi_{\mu\rho,\nu\sigma}^{\rm{non-minimal}}$ from the non-minimal
scalar one-loop (\ref{eq:Pi_nonminimal}) has the $Q_{\mu\rho,\nu\sigma}$
term that depends on the value of a (non-minimal) coupling constant
$\lambda$. 
Hence we can tune the coupling constant  
$\lambda$ so that the sum of $Q_{\mu\rho,\nu\sigma}$ terms from non-unitary
matters and from unitary non-minimal scalars may possess the 
minus sign relative to that of $P^{(2)}_{\mu\rho,\nu\sigma}$ terms. 
As an example, we first integrate both $N_{\rm{s}}$ unitary
non-minimal scalars and $N$ non-unitary matters defined in 
(\ref{eq:NUmatter}) (or $N$ U(1) FP ghost fields). 
Then, according to the degrees-of-freedom rule, 
the dressed graviton propagator turns out to have no spin-2 
tachyons if 
\begin{equation}
r\equiv \frac{N}{N_{\rm{s}}}> \frac{1}{2}.\label{eq:5.2}
\end{equation} 
Moreover, if we choose 
\begin{equation}
\lambda\leq \frac{-1-\sqrt{2r}}{16}~~\mbox{\rm or}~~
\lambda\geq \frac{-1+\sqrt{2r}}{16},\label{eq:5.3}
\end{equation}
there are no tachyons in the gauge part, either. 
Eqs. (\ref{eq:5.2}) and (\ref{eq:5.3}) are not 
severe restrictions. One may also include fermions with 
an appropriate
modification of the constraint for $\lambda$.

\section{Summary and Prospect}

\indent

 We have presented preliminary results for our work that
aims to get insight into universal properties of three- and 
four-dimensional quantum gravity coupled to matter fields.
Our general spirit in the studies of quantum gravity is the
following modest one.
Although the description of full quantum properties such as of the
unitarity problem may require a final theory such as superstring 
theories, it is reasonably expected that low-energy properties 
such as the phase structure or the stability of 
space-time vacuum and even some universal behaviors 
near a scaling region, if any, may be well described 
by {\rm renormalizable field theories}.
In the continuum approach, renormalizability is technically
important to make actual predictions especially for gauge
theories like gravity, where it is difficult to introduce
a cut-off scale consistently. Use of the $1/N$ expansion in 
quantum gravity is one possible direction to realize 
renormalizability by making much of matter fluctuations 
(back reaction) rather than space-time fluctuations.
It could also allow non-perturbative predictions such as 
a non-trivial fixed point in renormalization-group, and is 
thus worth studying.

As a preliminary step we have reported some results for
the applicability of the $1/N$ expansion to three-dimensional 
gravity. Continuing Kugo's work, we have first confirmed the 
generality as to the presence of tachyon in the spin-2 part of 
the graviton propagator dressed by {\em unitary} matters, and 
observed that the spin-2 part of matter one-loop corrections are 
quantized to $\kappa^2(-p^2)^{\frac{3}{2}}/512$ times the physical 
degrees of freedom. This ``degrees-of-freedom rule'' for matter 
one-loop corrections is an analogue of the one for the conformal 
anomaly in two dimensions. The graviton called here is the 
fluctuation around the flat space-time and the result implies 
that the flat space-time is quantum-mechanically unstable for 
large $N$ and higher-order analyses cannot work. In other words, 
in three-dimensional gravity coupled to unitary matters,
the $N \rightarrow \infty$ limit of matter fields is not a stable
zero-th order approximation for the flat space-time. The $\beta$ 
function for $\kappa$ exhibits no ultraviolet fixed point
in these cases, either.

To stabilize the flat space-time in the $1/N$ approach, 
the results then require the modification of the theory. 
We have suggested two possible cures
(the higher-derivative gravity and the coupling to non-unitary 
matters) by which a tachyon may be circumvented in the spin-2 
part of a graviton propagator. For unitary matters, it seems 
quite natural and appealing to introduce higher-derivative terms 
and to investigate further the renormalization effects from the 
``dressed'' graviton-loop appearing in the next-to-leading order.

From the purely theoretical viewpoint, 
the simplest and interesting case
may be Einstein gravity coupled to {\em non-unitary} matters to which
the $1/N$ expansion can be applied without introducing any new
parameters for the gravity part. 
Although being an unrealistic model, it allows a manifestly 
gauge-invariant
expansion keeping the renormalizability of {\em Einstein} gravity
and does not have a tachyonic pole at least in the spin-2 part 
of the graviton propagator.
Another fascinating point is that the theory is
expected to possess an ultraviolet fixed point for $\kappa$ 
even to leading order \cite{MY2}, which means that it could
serve as an interesting theoretical laboratory for studying 
analytically the critical behavior of gravitation.
In this theory, however, a tachyonic pole exists in the
gauge-dependent (spin-1 or -0) part of the graviton propagator
and we have verified that it can never be evaded by any choice 
in the class of linear gauges.
To cure this difficulty we are led to include 
unitary non-minimal scalars in addition.
Under the conditions (\ref{eq:5.2}) and (\ref{eq:5.3}),
the full theory is thus completely free from the tachyon disaster;
we can have a tractable quantum gravity model in which the flat
space-time is stable at $N\rightarrow\infty$.
For example, we can consider the following model lagrangian:
\begin{eqnarray}
{\cal L}&=&\frac{1}{\kappa^2}\sqrt{g}R~+\Lambda\sqrt{g}
     +~{\cal L}_{\rm{FP+GF}}  \nonumber\\
&&+~\sum_{i=1}^{N}\sqrt{g}g^{\mu\nu}\left(\partial_{\mu}\bar{c}_i
\partial_{\nu}c_i ~+~\frac{1}{2}
\partial_{\mu}\phi_i\partial_{\nu}\phi_i+ \lambda R\phi_i^2\right),
\label{eq:nulag}
\end{eqnarray} 
where $\bar{c}_i$,$c_i$ and $\phi_i$ are $N$-plet (non-unitary) 
anti-commuting scalar fields and (unitary) non-minimal scalar fields 
respectively and the strength of the non-minimal coupling $\lambda$ 
(not renormalized to leading order) is assumed to satisfy 
(\ref{eq:5.3}). 
This model, although unrealistic, contains only 
the minimal parameter 
$\kappa$ for self-gravity dynamics and allows a renormalizable
tachyon-free $1/N$ expansion. 
Further, the model is anticipated to possess an ultraviolet
fixed point for $\kappa$ and will serve as an fascinating statistical
model possessing two phases of space-time; to the leading order 
the transition is driven solely by matter fluctuations \cite{MY2}
and may also be confirmed by numerical simulations of models with 
many kinds of non-unitary matters. 
Further, this model allows the second-order calculation of
renormalization-group functions, where we may now consistently 
take the graviton-loop effects into account in three dimensions. 
The detailed analyses will appear elsewhere.
We hope that the results in this paper provide useful bases
that can be developed in several directions.

\vskip 10mm
\section*{acknowledgements}

        A part of the present work was carried out during our stay 
at Uji Research Center, Yukawa Institute for Theoretical Physics. 
We gratefully acknowledge the kind hospitality we enjoyed there.

        We would like to thank M. Ninomiya for useful discussions,
comments and reading the manuscript.
We are also grateful to T. Kugo, S. Sawada and S. Uehara for 
discussions.  The work of S.M. is supported by Soryuushi-Shougakukai 
and the Alexander von Humboldt Foundation.

\section*{Appendix A}

\indent

        In this appendix we describe the definitions 
of the projection
operators in the space of symmetric
rank-two tensors in three-dimensional spacetime \cite{N}. 
The projection operators 
in the spaces of spin-2, -1 and -0 are given by 
\begin{eqnarray}
P^{(2)}_{\mu\rho,\nu\sigma}
&=& \frac{1}{2}(\theta_{\mu\nu}\theta_{\rho\sigma}
+\theta_{\mu\sigma}\theta_{\nu\rho}-\theta_{\mu\rho}\theta_{\nu\sigma}),\\
P^{(1)}_{\mu\rho,\nu\sigma}&=& \frac{1}{2}
(\theta_{\mu\nu}\omega_{\rho\sigma}
+\theta_{\rho\sigma}\omega_{\mu\nu}
+\theta_{\rho\nu}\omega_{\mu\sigma}
+\theta_{\mu\sigma}\omega_{\rho\nu}),\\
P^{\rm{(0-s)}}_{\mu\rho,\nu\sigma}&=& \frac{1}{2}
\theta_{\mu\rho}\theta_{\nu\sigma},\\
P^{\rm{(0-w)}}_{\mu\rho,\nu\sigma}&=&
\omega_{\mu\rho}\omega_{\nu\sigma},
\end{eqnarray}
where
\begin{equation}
\theta_{\mu\nu}=\eta_{\mu\nu}- ~\frac{p_{\mu}p_{\nu}}{p^2},
~~\omega_{\mu\nu}= \frac{p_{\mu}p_{\nu}}{p^2}.
\end{equation}
They satisfy the completeness condition 
\begin{equation}
P^{(2)}_{\mu\rho,\nu\sigma}
+P^{(1)}_{\mu\rho,\nu\sigma}
+P^{\rm{(0-s)}}_{\mu\rho,\nu\sigma}
+P^{\rm{(0-w)}}_{\mu\rho,\nu\sigma}=1.
\end{equation}
The tensors which intertwine the two spin-0 subspaces are 
\begin{eqnarray}
P^{\rm{(0-sw)}}_{\mu\rho,\nu\sigma}&=& \frac{1}{\sqrt{2}}
\theta_{\mu\rho}\omega_{\nu\sigma},\\
P^{\rm{(0-ws)}}_{\mu\rho,\nu\sigma}&=& \frac{1}{\sqrt{2}}
\omega_{\mu\rho}\theta_{\nu\sigma}.
\end{eqnarray}
The tensor of spin-0 and -1 parts in the graviton two-point function 
is given by  
\begin{equation}
Q_{\mu\rho,\nu\sigma}=
P^{\rm{(0-s)}}_{\mu\rho,\nu\sigma}
+\sqrt{2}(P^{\rm{(0-sw)}}_{\mu\rho,\nu\sigma}
+P^{\rm{(0-ws)}}_{\mu\rho,\nu\sigma}) 
+2P^{\rm{(0-w)}}_{\mu\rho,\nu\sigma}.
\end{equation}
It is convenient to represent the tensor algebra by $4\times 4$
matrices. Let $r$ be a linear map such that 
\begin{eqnarray}
&&r\left(AP^{(2)}_{\mu\rho,\nu\sigma}
+BP^{(1)}_{\mu\rho,\nu\sigma}
+CP^{\rm{(0-s)}}_{\mu\rho,\nu\sigma}
+D(P^{\rm{(0-sw)}}_{\mu\rho,\nu\sigma}
   +P^{\rm{(0-ws)}}_{\mu\rho,\nu\sigma})
+EP^{\rm{(0-w)}}_{\mu\rho,\nu\sigma}\right)\nonumber\\
&&\nonumber\\
&=&\left[
\begin{array}{cccc}
A&&&\\
&B&&\\
&&C&D\\
&&D&E
\end{array}
\right].
\end{eqnarray} 
Then if $r(T_{\mu\rho,\nu\sigma})=M$ and $r(T'_{\mu\rho,\nu\sigma})=M'$, 
$r(T_{\mu\rho,\alpha\beta}T'^{\alpha\beta,}_{~~~~\nu\sigma})$ 
is equal to 
$MM'$. In other words, $r$ is a representation of this algebra.

\section*{Appendix B}

\indent

        In this appendix we summarize some formulas for the
transformations from the bilinear forms in $\widetilde{f}_a^{~\mu}$
(\ref{eq:tf}) and
$h_{\mu\nu}$ (\ref{eq:h->th}) to that in $\widetilde{h}^{\mu\nu}$. 
Let 
\begin{equation}
U^{\mu\rho\nu\sigma}_{~~~~~\alpha\beta\gamma\delta}(x,y)
=
(x\eta_{\alpha\beta}\eta^{\mu\rho}
+y\delta_{\alpha}^{\mu}\delta_{\beta}^{\rho})
(x\eta_{\gamma\delta}\eta^{\nu\sigma}
+y\delta_{\gamma}^{\nu}\delta_{\delta}^{\sigma}).
\end{equation}
$(x,y)=(\frac{1}{2},\frac
{1}{2})$ and $(1,-1)$ 
correspond to the transformations 
$\widetilde{f}_a^{~\mu}\rightarrow\widetilde{h}^{\mu\nu}$ and
$h_{\mu\nu}\rightarrow\widetilde{h}^{\mu\nu}$, respectively.
Then
\begin{eqnarray}
&&\eta_{\mu\nu}\eta_{\rho\sigma}
U^{\mu\rho\nu\sigma}_{~~~~~\alpha\beta\gamma\delta}(x,y)\nonumber\\
&=&
(nx^2+2xy)\eta_{\alpha\beta}\eta_{\gamma\delta}
+y^2\eta_{\alpha\gamma}\eta_{\beta\delta},\\
&&\nonumber\\
&&\eta_{\mu\rho}\eta_{\nu\sigma}
U^{\mu\rho\nu\sigma}_{~~~~~\alpha\beta\gamma\delta}(x,y)\nonumber\\
&=&(nx+y)^2\eta_{\alpha\beta}\eta_{\gamma\delta},\\
&&\nonumber\\
&&\eta_{\mu\nu}p_{\rho}p_{\sigma}
U^{\mu\rho\nu\sigma}_{~~~~~\alpha\beta\gamma\delta}(x,y)\nonumber\\
&=&
x^2p^2\eta_{\alpha\beta}\eta_{\gamma\delta}
+xy(\eta_{\alpha\beta}p_{\gamma}p_{\delta}+
\eta_{\gamma\delta}p_{\alpha}p_{\beta})
+y^2\eta_{\alpha\gamma}p_{\beta}p_{\delta},\\
&&\nonumber\\
&&\eta_{\mu\rho}p_{\nu}p_{\sigma}
U^{\mu\rho\nu\sigma}_{~~~~~\alpha\beta\gamma\delta}(x,y)\nonumber\\
&=&
(nx+y)xp^2\eta_{\alpha\beta}\eta_{\gamma\delta}
+(nx+y)y\eta_{\alpha\beta}p_{\gamma}p_{\delta},\\
&&\nonumber\\
&&
p_{\mu}p_{\nu}p_{\rho}p_{\sigma}
U^{\mu\rho\nu\sigma}_{~~~~~\alpha\beta\gamma\delta}(x,y)\nonumber\\
&=&
x^2(p^2)^2\eta_{\alpha\beta}\eta_{\gamma\delta}
+xyp^2
(\eta_{\alpha\beta}p_{\gamma}p_{\delta}+
\eta_{\gamma\delta}p_{\alpha}p_{\beta})
+y^2p_{\alpha}p_{\beta}p_{\gamma}p_{\delta},
\end{eqnarray}
where $n$ is the dimensions of spacetime ($n=3$).

\newpage

\section*{Figure captions}
\noindent
Figure 1:~(a)The dressed graviton propagator.
(b)The dressed $N$-point vertex.\\
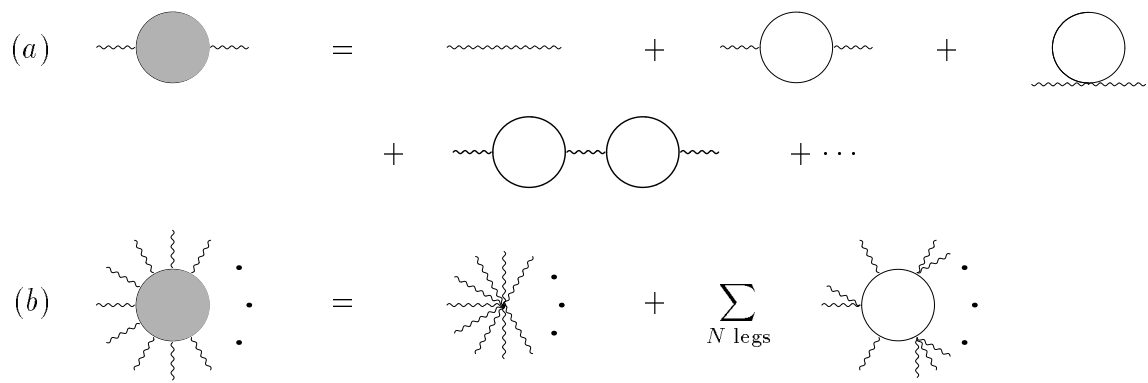
Figure 2:~The scalar one-loop diagrams.
Bold and wavy lines stand for scalar 
and graviton, repsectively.\\
Figure 3:~The spinor one-loop vacuum 
polarization.\\
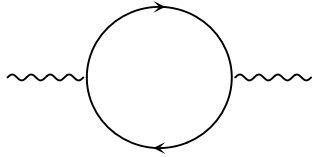
Figure 4:~The U(1) gauge boson (curly 
lines) and U(1) FP ghost (broken lines)
one-loop vacuum polarizations.\\
Table 1:~The Feyman rules. The momenta 
of the vertices are taken in-going.

\begin{thebibliography}{99}
\bibitem{DTL}
D. V. Boulatov and A. Krzywicki, Phys. Lett. {\bf B266} (1991) 285\\
J. Ambj\o rn and S. Varsted, Nucl. Phys. {\bf B373} (1992) 557\\
M. E. Agishtein and A. A. Migdal, Nucl. Phys. {\bf B385} (1992) 395
\bibitem{KKN}H. Kawai and M. Ninomiya, Nucl. Phys. {\bf B336} (1990)
115\\
H. Kawai, Y. Kitazawa and M. Ninomiya, Nucl. Phys. {\bf B393} (1993)
280; {\bf B404} (1993) 684
\bibitem{CSW}E. Witten, Nucl. Phys. {\bf B311} (1988) 46
\bibitem{Ash}A. Ashtekar, Phys. Rev. Lett. {\bf 57} (1986) 2244; 
Phys. Rev. {\bf D36} (1987) 1587
\bibitem{Weinberg}S. Weinberg, in: {\it``General Relativity''}, 
an Einstein Centenary Survey, 
eds. S. W. Hawking and W. Israel 
(Cambridge University Press 1979) 790, and references therein
\bibitem{4fermi}
G. Parisi, Nucl. Phys. {\bf B100} (1975) 368\\
D. J. Gross, in {\it``Methods in Field Theory''},
Les Houches Summer School in Theoretical Physics, 
eds. R. Balian and J. Zinn-Justin 
(North-Holland, Amsterdam 1976)\\
T. Eguchi, Phys. Rev.  {\bf D14} (1976) 2755\\
K. Shizuya, Phys. Rev.  {\bf D24} (1980) 2327\\
B. Rosenstein, B. J. Warr and S. H. Park,  Phys. Rev. Lett.
{\bf 62} (1989) 1433
\bibitem{NLSM}
I. Ya. Aref'eva, Teor. Mat. Fiz. {\bf 29} (1976) 147;
{\bf 31} (1977) 3;
{\bf 36} (1978) 24;
{\bf 36} (1978) 159;
{\bf 117} (1979) 393\\
I. Ya. Arefe'va and S. I. Azakov,  Nucl. Phys. {\bf B162} (1980) 298
\bibitem{MSHKK} E. Manousakis and R. Salvator, 
Phys. Rev. Lett. {\bf 62}
(1989) 1310; Phys. Rev. {\bf B40} (1989) 2205\\
S. Hands, A. Koci\'{c }and J.B. Kogut, preprint CERN-TH-6557-92 /
ILL-(TH)-92-\#19
\bibitem{T}
E. T. Tomboulis, Phys. Lett. {\bf 70B} (1977) 361;
{\bf 97B} (1980) 77
\bibitem{AT}I. Antoniadis and E. T. Tomboulis, Phys. Rev. {\bf D33}
(1986) 2757
\bibitem{Johnston}D. A. Johnston, Nucl. Phys. {\bf B297} (1988) 721
\bibitem{Kugo}T. Kugo, {\it``Renormalizability of Three-Dimensional
Gravity Coupled to \\Matter''}, Kyoto preprint KUNS1014 HE(TH)90/05,
April (1990)
\bibitem{Weinbergtheorem}S. Weinberg, Phys. Rev. {\bf 118} (1960) 838
\bibitem{phi4}S. Coleman, R Jackiw and H. D. Politzer, 
Phys. Rev. {\bf
D10} (1974) 2491
\bibitem{AP}T. Appelquist and R. D. Pisarski, Phys. Rev. {\bf D23}
(1980) 2305
\bibitem{gravitationalCS1}J. J. van der Bij, 
R. D. Pisarski and S. Rao
Phys. Lett. {\bf B179} (1986) 87
\bibitem{gravitationalCS2}S. Deser, R. Jackiw and S. Templeton, 
Phys. Rev. Lett. {\bf 48} (1982) 975; Ann. Phys. {\bf 140} (1982)
372\\
S. Deser and Z. Yang, Class. Quantum Grav. {\bf 7} (1990) 1603
\bibitem{Pi1}C. Bernard and A. Dunkan, Ann. Phys. 
{\bf 107} (1977) 201
\bibitem{Pi2}M. A. Go\~{n}i and M. A. Valle, Phys. Rev. {\bf D34} 
(1986) 648
\bibitem{Kugo2}T. Kugo, private communication
\bibitem{KugoUehara}T. Kugo and S. Uehara, 
Nucl. Phys. {\bf B197} (1982) 378
\bibitem{DNOPZ}B. Durhuus, H. B. Nielsen, P. Olesen and E. Peterson, 
Nucl. Phys. {\bf B196} (1982) 498\\
A. B. Zamolodchikov, Phys. Lett. {\bf 117B} (1982) 87
\bibitem{KPZDDK}
V.G. Knizhnik, A.M. Polyakov and A.A. Zamolodchikov, Mod. Phys. Lett.
{\bf A3} (1988) 819;\\
F. David, Mod. Phys. Lett. {\bf A3} (1988) 651;\\
J. Distler and H. Kawai, Nucl. Phys. {\bf B321} (1989) 509
\bibitem{KKSW}
N. Kawamoto, V.A. Kazakov, Y. Saeki and Y. Watabiki,
Phys. Rev. Lett. {\bf 68} (1992) 2113
\bibitem{MY2} S. Mizoguchi and H. Yamamoto, in preparation
\bibitem{N}P. van Nieuwenhuizen, Nucl. Phys. {\bf B60} (1973) 478
\end{thebibliography}
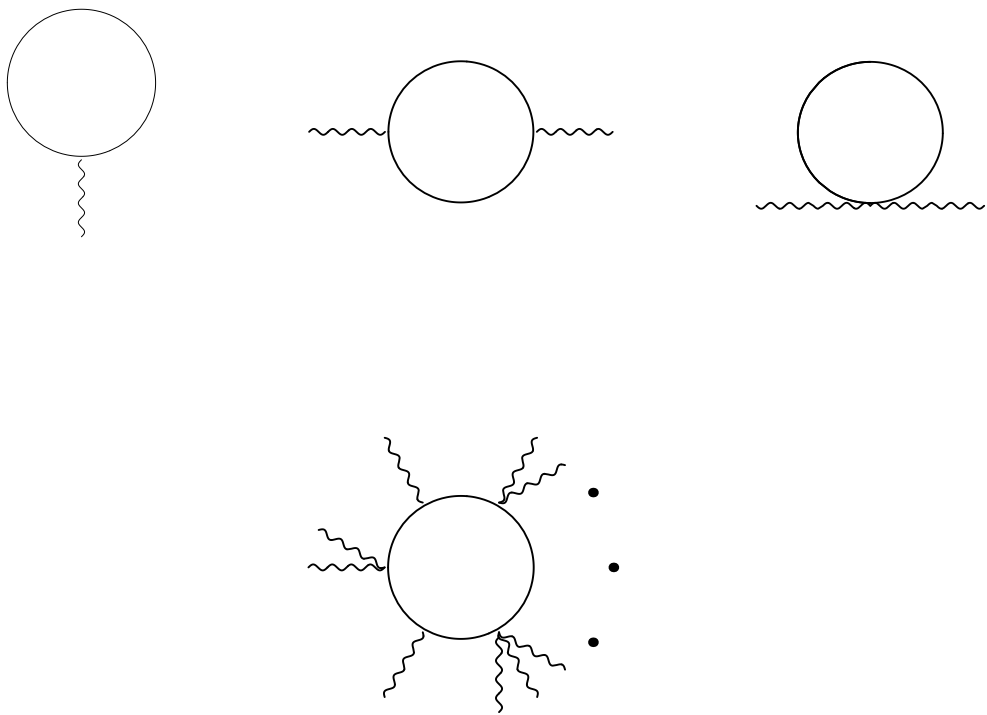
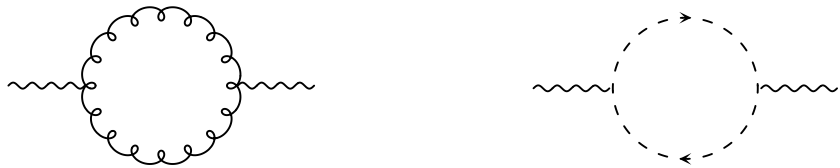
\end{document}